\documentclass[preprint]{vgtc}                          

\graphicspath{{figures/}{pictures/}{images/}{./}} 

\usepackage{times}                     

\usepackage{tabu}                      
\usepackage{booktabs}                  
\usepackage{lipsum}                    
\usepackage{mwe}                       

\usepackage{mathptmx}                  

\onlineid{0}

\vgtccategory{Research}

\vgtcinsertpkg

\title{Counterpoint: Orchestrating Large-Scale Custom Animated Visualizations}

\author{%
  Venkatesh Sivaraman\thanks{e-mail: \{venkats, fje, domoritz, adamperer\}@cmu.edu} \and
  Frank Elavsky$^\ast$ \and
  Dominik Moritz$^\ast$ \and Adam Perer$^\ast$\\%
  \parbox{6.5in}{\scriptsize \centering Carnegie Mellon University}
}

\abstract{%
  Custom animated visualizations of large, complex datasets are helpful across many domains, but they are hard to develop. Much of the difficulty arises from maintaining visualization state across many animated graphical elements that may change in number over time. We contribute Counterpoint, a framework for state management designed to help implement such visualizations in JavaScript. Using Counterpoint, developers can manipulate large collections of marks with reactive attributes that are easy to render in scalable APIs such as Canvas and WebGL. Counterpoint also helps orchestrate the entry and exit of graphical elements using the concept of a rendering ``stage.'' Through a performance evaluation, we show that Counterpoint adds minimal overhead over current high-performance rendering techniques while simplifying implementation. We provide two examples of visualizations created using Counterpoint that illustrate its flexibility and compatibility with other visualization toolkits as well as considerations for users with disabilities. Counterpoint is open-source and available at \url{https://github.com/cmudig/counterpoint}.
}

\keywords{Visualization Toolkits, Animation, Web Interfaces, Software System Structures.}

\graphicspath{{figs/}{figures/}{pictures/}{images/}{./}} 

\usepackage{enumitem}

\usepackage[svgnames]{xcolor}
\usepackage[frozencache,cachedir=.]{minted2}

\begin{document}


\firstsection{Introduction}

\maketitle

Web-based interactive data visualizations allow users to consume and explore large quantities of data across diverse computing environments. 
Toolkits such as D3~\cite{bostock_d_2011} and Vega~\cite{satyanarayan2015reactive,satyanarayan_vega-lite_2017} have greatly accelerated the creation of these online data visualizations, helping software developers perform common tasks such as building statistical graphics, laying out network diagrams, and handling interaction paradigms such as brushing-and-linking. 
While these tools have been sufficient to create many sophisticated and useful visualizations, it remains difficult to venture beyond these tools’ standard functionality in terms of scalability, animation, or accessibility. 
For example, many tools for machine learning (ML) aim to visualize model embeddings as scatter plots with thousands to millions of points~\cite{sivaraman_emblaze_2022,Smilkov2016}, but implementations of such scatter plots using these toolkits tend to lag when interactions and animations are added. 
Building a smoothly-animating scatter plot from scratch often results in complex code that is harder to modify later and a chart that is less responsive to end-user accessibility requirements, such as interoperability with assistive technologies.

These limitations arise from the trade-offs inherent in current Web graphics APIs, including Scalable Vector Graphics (SVG), HTML5 Canvas, and WebGL. 
Using native SVG, for instance, developers can take advantage of event handling and accessibility features built into HTML and CSS; however, rapidly modifying SVG elements results in expensive redraw operations and lowered framerates. 
Conversely, Canvas and WebGL can support smooth rendering at large scales, but they do not directly provide affordances for interaction, animation, and accessibility. 
As a result, the cost of this flexibility and performance is often a longer development time and more complex code tied to the rendering framework.

We propose to improve the scalability-complexity trade-off in Web graphics APIs by designing rendering-agnostic data structures for visualization state. 
A central challenge in software development, \textit{state management} concerns how the state of a user interface is stored, read, and manipulated as a system is used.
Most modern reactive web frameworks include some form of state management, such as React/Redux or Svelte Stores.
However, these systems do not translate well to creating visualizations, which require tracking the state of large time-varying collections of marks with animated properties.
Tools for state management designed specifically for large-scale data visualizations could help developers orchestrate changes across different interactions and animations in the rendering environment they prefer. 

We contribute Counterpoint, a state management library that makes it easier to create custom, scalable, animated visualizations. 
Counterpoint provides data types to represent animatable attributes, marks that contain attribute collections, and render groups that manage a potentially time-varying set of marks with independent animations. 
Developers can use these objects to represent their visualization's state, then query their values to simplify rendering code. 
Furthermore, Counterpoint provides utilities on top of its primary types to facilitate interaction tasks such as zooming and hit-testing. 
Through a performance benchmark, we demonstrate the potential for Counterpoint to accelerate visualizations to Canvas speeds while maintaining the code simplicity of SVG. 
Then we present two case studies that demonstrate Counterpoint’s ability to easily express animations and work in tandem with other visualization frameworks.

\section{Related Work}


\subsection{Animation in Large-Scale Data Visualization}

More than a simple technique to engage viewers, animation is an essential aspect of visualization design that can help communicate a desired message or reveal useful insights~\cite{steele_beautiful_2010}.
In recent years Web-based animated visualizations have been used to communicate important and complex data to the public, ranging from global health trends~\cite{gapminder_gapminder_nodate} to COVID-19 infographics~\cite{samet_using_2020} to interactive election forecasts~\cite{noauthor_election_2020,bycoffe_what_2015}.
In these instances, animation effectively conveys change over time, relative magnitude, and uncertainty.
While existing visualization toolkits can help developers create and render these graphics, Counterpoint is designed to help manage complexity in their implementation as the number of animations grows.

An important set of use cases for large-scale animated plots is the visualization of specialized datasets for experts. 
For example, in machine learning, animations in scatter plots of vector embeddings help model developers understand what neural networks have learned and when they might err~\cite{Smilkov2016,robertson_angler_2023,sivaraman_emblaze_2022}. 
Other systems use interactive animations to help students and developers understand how neural networks transform data~\cite{wang_cnn_2021,yeh_attentionviz_2023}.
In computational biology, researchers use large scatter plots and network diagrams to understand relationships between genes, proteins, and cell types~\cite{lyi_gosling_2022,zhou_networkanalyst_2019,zhou_omicsnet_2018}.
Our work aims to reduce the technical burden of developing scalable animations in these bespoke visualizations.

\subsection{Web Frameworks for Visualization}

Several tools exist to help developers create data visualizations on the Web, but state management remains an under-addressed challenge.
Most prominently, D3~\cite{bostock_d_2011} helps developers create SVG-based data visualizations by storing state directly in the Document Object Model (DOM), which tends to work best for non-reactive properties and is difficult to extend to draw loop-based rendering frameworks.
General-purpose graphics libraries such as Pixi.js~\cite{noauthor_pixijs_nodate}, regl~\cite{noauthor_regl_nodate}, and Two.js~\cite{noauthor_twojs_nodate} can help developers attain higher-performance rendering, but they do not assist with state management and reactivity.
Meanwhile, higher-level APIs for data visualization such as Vega-Lite~\cite{satyanarayan_vega-lite_2017} and ECharts~\cite{li_echarts_2018} require little code to create animated charts, but it can be difficult to render custom visual elements or transitions.

Existing tools for state management tend to be focused on communicating state across components of an application, not on orchestrating interaction types within a single view.
For example, reactive frameworks such as React, Vue, and Svelte typically use a companion framework for state management (Redux, Vuex, or Svelte stores) to store global state.
Specific to visualization, Mosaic~\cite{heer_mosaic_2023} provides a ``middle-tier'' architecture that in essence provides state management for multiple coordinated views.
While these approaches work well for non-animated interfaces with a relatively small number of state elements, our work helps manage visualization state that may not only be reactive and animated, but also distributed over a large, time-varying array of graphical elements.

\subsection{Animation Accessibility in Visualization}
Accessibility and disability-focused design in data visualization is an emerging area in both research and practice~\cite{Kim2021, Hsueh2023, Wimer2024}. 
Animation affects both screen reader users (most commonly users who are blind), whose technology limits their ability to perceive live spatial change in elements, and users with motion/vestibular and photosensitive disabilities, who may experience adverse effects (such as dizziness, vertigo, or seizures) when viewing animation~\cite{South2022}. 
However, while accessibility standards and work in HCI acknowledge the impact of animated visualization on these users, it remains difficult to effectively accommodate their needs with existing toolkits~\cite{Kim2023,Elavsky2022,Fan2023}. 
By helping developers maintain state across animations, Counterpoint can help them better address these concerns.

\section{Counterpoint Library}

Counterpoint is a TypeScript library that defines useful types for state management in large-scale animated visualizations.
We built Counterpoint through an iterative process that took place alongside the development of two major interactive visualization tools~\cite{sivaraman_emblaze_2022,boggust_cnc_2024}, as well as several smaller demo projects.
While we were able to use frameworks such as D3, regl, and Svelte to quickly create initial prototypes, the need for a robust tool for state management became apparent as the number of interactions and animations grew.
For example, clicking a point in the scatter plot in Emblaze~\cite{sivaraman_emblaze_2022} could lead to changes in points' opacity and radius, the appearance of some connecting lines and the disappearance of others, as well as changes to the zoom transform.
Animations played a key role in making these visual changes legible to the user, yet they became prohibitively complicated to implement in combination.
We therefore developed the following goals for Counterpoint:
\begin{enumerate}[label={\bfseries G\arabic*.}, ref={\bfseries G\arabic*},itemsep=1ex, labelindent=0pt, wide=0pt, labelwidth=!]
    \item \textit{Help developers manage code complexity by balancing declarative and imperative approaches to state management.} Analogous to reactive web frameworks, Counterpoint allows developers to easily specify visualization state through declarative, dynamically-computed attributes while still permitting imperative approaches when needed for simpler or faster code. \label{goal:declarative-imperative}
    \item \textit{Facilitate efficient draw loop-based rendering across frameworks.} The draw loop is common to most scalable graphics engines, so we designed Counterpoint's reactivity and animation logic to work efficiently with a ticker that runs at every frame. \label{goal:draw-loop-rendering}
    \item \textit{Expose representations of both intermediate and final visualization states.} Animations rendered using a custom draw loop require access to attribute values at every frame, posing a conceptual issue of how to distinguish between the final \textit{specified} state of the visualization and the \textit{momentary} states that arise during animations. We opted to make both types of state available in Counterpoint. \label{goal:intermediate-and-final-state}
\end{enumerate}

\subsection{Marks, Attributes, and Render Groups}

The basic unit of a visualization created with Counterpoint is a \texttt{Mark}, referring to a graphical element that represents data in some way ~\cite{munzner_visualization_2014}. 
Marks can represent any visual element with a set of potentially time-varying characteristics, each represented by a member \texttt{Attribute} object (described below).
For example, a point mark could be represented using Attributes such as \texttt{x}, \texttt{y}, \texttt{radius}, and \texttt{color}.
Marks can be associated with a data object using the \texttt{represented} property, allowing developers to easily keep track of which marks are associated with which underlying data elements.

An attribute's value can either be set directly or computed using a callback function at every frame, allowing users the option of using a declarative or imperative approach (\ref{goal:declarative-imperative}).
For example, in the below snippet the resulting mark has two numerical Attributes representing its position and one string Attribute that changes dynamically as the $x$-coordinate changes:
\begin{minted}{javascript}
let mark = new Mark('mark-id', {
  x: 250, y: 400, 
  color: () => `hsl(${mark.attr('x') % 360}, 60%, 60%)`
}
\end{minted}
The \mintinline{javascript}{animate()} or \mintinline{javascript}{animateTo()} methods of a mark or render group execute animations between a current attribute value and a new computed or static value, respectively. 
In addition to configuring the duration, easing function, and delay for any animation, developers can specify a custom interpolator that should be used to map the animation progress to an intermediate value, such as those provided by \texttt{d3-interpolate}.
Addressing \ref{goal:intermediate-and-final-state}, attributes expose both momentary and specified states through different accessors, enabling computations that work independently of animations.


To limit data updates and redraws wherever possible (\ref{goal:draw-loop-rendering}), most Counterpoint types share an \mintinline{javascript}{advance()} method which synchronizes time across the objects and determines whether the element has changed since the last \mintinline{javascript}{advance()} call.
A redraw will only occur when at least one attribute's \mintinline{javascript}{advance()} method returns \mintinline{javascript}{true}.
To further improve efficiency, render groups track changes to the marks they contain through internal listener callbacks, so they can efficiently detect which marks need to be advanced at each frame.

Consistent with Counterpoint's focus on state management alone, none of these objects make assumptions on how they will be rendered; marks and attributes can represent any types of time-varying properties that the developer would like to track. 

\subsection{Choreographing Mark Entry and Exit}

A frequent challenge when implementing interactive visualizations is managing the animated entry and exit of marks.
For example, while D3's \texttt{transition} operator can be applied to \texttt{enter} and \texttt{exit} selections, it can be difficult to specify how elements should behave when they need to re-enter while they are exiting or vice versa.
Therefore, Counterpoint includes a system for \textbf{staging} marks that explicitly accounts for interruptible entry and exit animations.

\begin{figure}
    \centering
    \includegraphics[width=0.9\linewidth, alt={Scatter plot of the cars dataset with two panels, one showing cars made before 1972 and one with more points showing cars made before 1980.}]{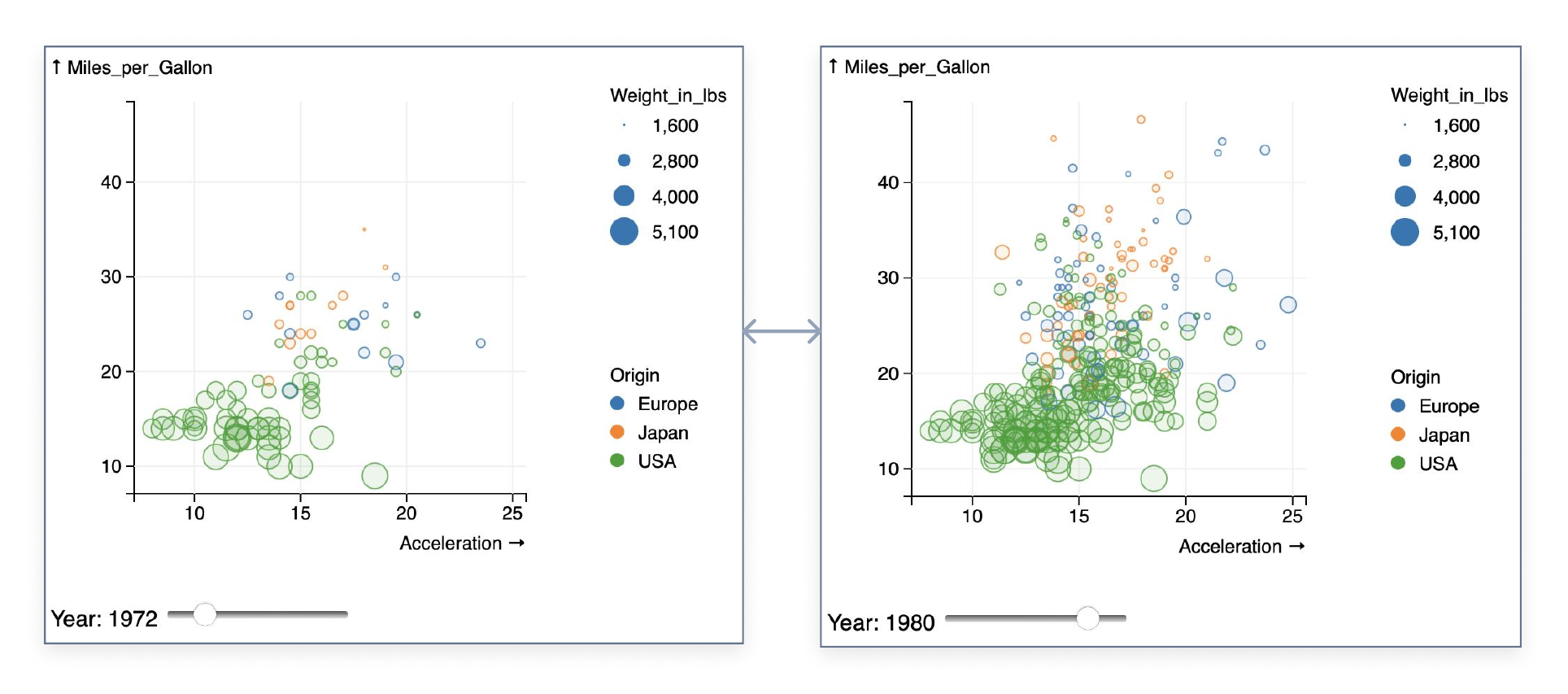}
    \caption{A visualization using the cars dataset~\cite{noauthor_vegavega-datasets_2024} which, when animated, demonstrates how staging helps differentiate the specified state (all cars produced before the current year) and the momentary state (marks that are entering and/or exiting based on the previous selection). The animated version is one of the examples at \url{https://dig.cmu.edu/counterpoint}.}
    \label{fig:cars-bubble}
\end{figure}

For a simple example, imagine we are building a scatter plot showing cars as they were manufactured over time (\cref{fig:cars-bubble}).
While a non-animated version of this visualization could easily be created using off-the-shelf tools like Vega-Lite~\cite{satyanarayan_vega-lite_2017}, we may want to make the chart more engaging by animating the alpha and radius when bubbles are added or removed as the user changes the year.
This can quickly become complicated to implement in existing frameworks because their APIs do not expose transient states, making it difficult to handle common situations such as moving the slider too quickly or receiving other interactions while animations are still playing.

To implement this chart using Counterpoint, we first create a \texttt{MarkRenderGroup} that will contain marks for all cars produced before the currently-selected year.
We configure the render group's staging behavior using up to three function callbacks:
\mintinline{javascript}{initialize(mark)}, to prepare a new \texttt{Mark} for animated entry (e.g. setting its alpha and radius to zero); \mintinline{javascript}{enter(mark)}, to animate the mark's entry (e.g. animating the alpha and radius up) and return a \mintinline{javascript}{Promise} that resolves when the animation is complete; and \mintinline{javascript}{exit(mark)}, to hide the mark and return a \mintinline{javascript}{Promise}.
To implement the draw loop, we draw all ``onstage'' marks by iterating over the \texttt{stage} property of the render group.
Addressing \ref{goal:intermediate-and-final-state}, staging allows us to specify visualization state on the render group while still being able to access all marks that are actually visible due to the momentary state. 


\subsection{Tools for Performance and Responsiveness}

Once a developer has implemented a visualization using Counterpoint's data types, they can take advantage of helpers that operate on Counterpoint structures to speed up useful visualization tasks:

\textbf{Spatial lookup for interaction.} 
While SVG can leverage JavaScript event listeners to respond directly to user interactions on an element, Canvas and WebGL APIs only detect interaction events at the level of the \mintinline{html}{<canvas>} element, not at individual marks.
Counterpoint's \texttt{PositionMap} class uses a spatial hashing approach to index mark locations, enabling constant-time lookup on average and linear-time creation of the hash table when coordinates change.

\textbf{Zooming and following marks.}
Counterpoint provides a \texttt{Scales} class to facilitate smooth user-initiated and programmatic zoom and pan.
These scales use \texttt{Attribute} objects to track the current zoom transform, so the scales can be smoothly animated just like marks.
Scales can also take advantage of attributes' reactivity to zoom to, center, or follow specific sets of marks.

\textbf{Responding to user preferences.}
Responding to users' system-wide accessibility settings, such as reduced motion or increased contrast, can help a broader range of people comfortably use visualizations. 
Counterpoint represents these settings in a \texttt{RenderContext} type which can be used like a reactive \texttt{Attribute}.
Developers can access these attribute values in the draw loop and update the rendering.

\textbf{Running animations in shaders.}
While \mintinline{javascript}{Attribute.get()} is designed for draw loops that are called on the CPU, additional optimization using WebGL can enable implementations that perform animation directly on the GPU.
Attributes can be designated as ``preloadable,'' to specify that when requested, they should return not the current value, but instead four values representing the initial and final values and times for the attribute's animation. 
These four values can be sent to the GPU at the start of an animation, so a GPU shader can compute the intermediate values at each frame using only the current time.

\section{Performance Evaluation}

\begin{figure}
        \centering
    \includegraphics[width=0.85\linewidth, alt={Line chart showing the time per frame in milliseconds for 6 implementations of the same scatter plot. For SVG, Counterpoint is slower than D3 for 1,000-10,000 points but faster for 50,000 and 100,000 (ranges between 16 and 607 ms per frame). For Canvas, Counterpoint is the same or slightly slower than D3 (ranges from 16 to 145 ms per frame). For regl, Counterpoint is faster at all plot sizes (ranges from 16 to 20 ms per frame).}]{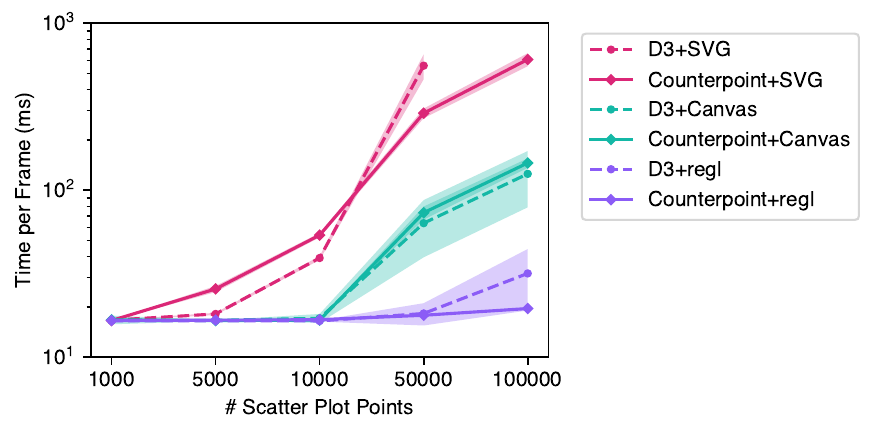}
    \caption{Average frame time for 5-second scatter plot animations using D3 and Counterpoint-based implementations. Shaded intervals represent 1 standard deviation around the mean over 20 trials.}
    \label{fig:performance-eval}
\end{figure}

We conducted a benchmark analysis to measure the performance overhead incurred by Counterpoint's state maintenance compared to commonly-used approaches.
We implemented six versions of the same animated scatter plot, using D3 and Counterpoint with SVG, Canvas, or regl~\cite{noauthor_regl_nodate} for rendering.\footnote{For D3+Canvas, we adapted the approach described in this tutorial: \url{https://www.bocoup.com/blog/smoothly-animate-thousands-of-points-with-html5-canvas-and-d3}.}
In each plot, we measured the duration of each frame of a 5-second animation in which each point moved to a new random location.
Additionally during each animation, 25\% of the points were removed from the plot using a fade-out transition while an equal number of new points were transitioned in. 
Evaluations were conducted on a MacBook Pro with an M2 Pro chip (16-core integrated GPU).

As shown in the chart of frame time results in \cref{fig:performance-eval}, Counterpoint's performance is comparable to or better than D3 across rendering frameworks, particularly in larger plots. 
The D3+SVG implementation scales poorly past 5,000 points, and it fails to load for the largest plot size. 
With Canvas rendering, the Counterpoint is almost identical to the D3 implementation for 10,000 points or fewer and within 1.2x for larger numbers of points. 
Note that the D3 implementation uses hardcoded animation logic on a shared timer for all marks, reducing the number of necessary function calls but making the code harder to modify.

While D3 requires taking a substantially different approach when switching from SVG to Canvas or regl, comparison of the implementations\footnote{\url{https://github.com/cmudig/counterpoint/tree/main/examples/performance_eval}} shows how the visualization state logic largely remains the same regardless of renderer.
Consistency across different rendering approaches could help developers quickly swap in a higher-performance renderer when deemed necessary, or even support multiple renderers for different browsing environments.


\section{Use Cases}

\subsection{Visualizing Citation Networks Over Time}

\begin{figure}
    \centering
    \includegraphics[width=0.8\linewidth, alt={Two screenshots of the citation scatterplot. In the first screenshot the full plot is visible, while in the second a paper called "Pixel bar charts" is selected and a second paper, "explAIner" is hovered to show the connection path between the two papers.}]{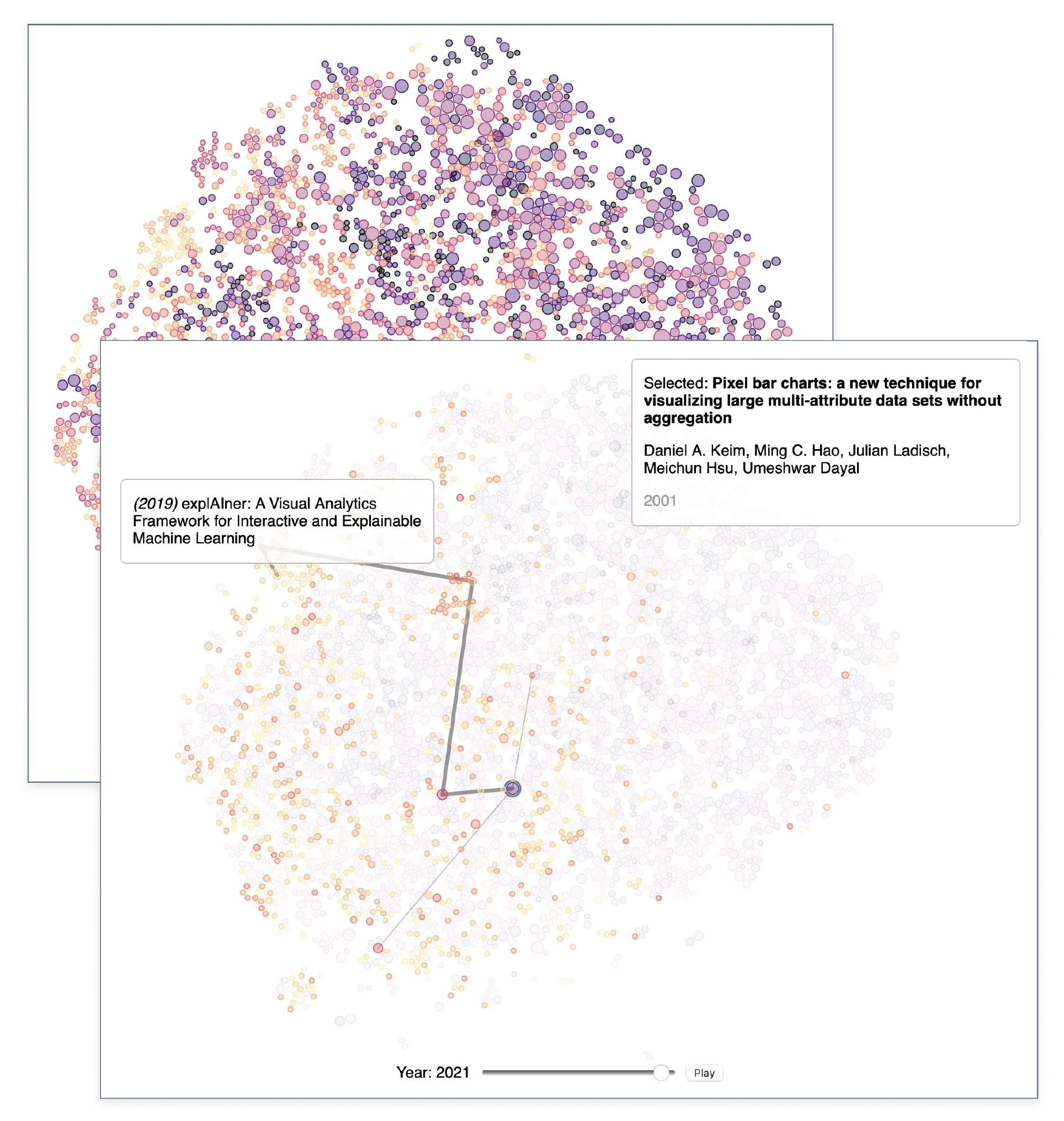}
    \caption{Visualizing citation networks and semantic similarity in IEEE VIS papers. Animations of mark opacity, connection lines, and reactive zoom help users trace lineage across papers. The interactive version is available at \url{https://dig.cmu.edu/counterpoint}.}
    \label{fig:citations}
\end{figure}

We demonstrate how Counterpoint enables large-scale visualizations with multiple coordinated animations using an example built on the Visualization Publications Dataset~\cite{Isenberg:2017:VMC}, shown in \cref{fig:citations}. 
To allow users to explore how visualization papers have influenced each other over time, we would like to visualize both semantic similarities and citation-based connections between papers.
Therefore, we compute a 2-dimensional layout of the 3,620 papers in the dataset using sentence embeddings of their titles.
Then, we construct a bubble chart where the color of each node corresponds to its year of publication and the size represents the number of citations, as shown in \cref{fig:citations} (left).
As the user changes the year, we use Counterpoint's stage mechanism to add publications to the plot that were released prior to the selected year.

While this gives us an overview of the dataset over time, interaction is necessary for deeper exploration of the papers.
Using Counterpoint's \texttt{PositionMap} we implement hover and select interactions that reveal edges between the current paper and any papers that cite it.
We also leverage \texttt{Scales} to zoom to the vicinity of the selection, and dynamically adjust the field of view as the year changes to show the paper's expanding sphere of influence.
Finally, to see at a glance how a hovered paper is related to the selected one, we add \texttt{Mark}s to represent citation paths in a second render group.

\subsection{Accessible Visualization of Global Economies}
\begin{figure}[t]
    \centering
    \includegraphics[width=0.9\linewidth, alt={Screenshot of a Gapminder plot with GDP per capita on the x axis, life expectancy on the y axis, and Serbia selected. A line shows Serbia's path through the plot over time and semitransparent bubbles show its position every 5 years. A text description provides a summary of the trend in the data over the years shown.}]{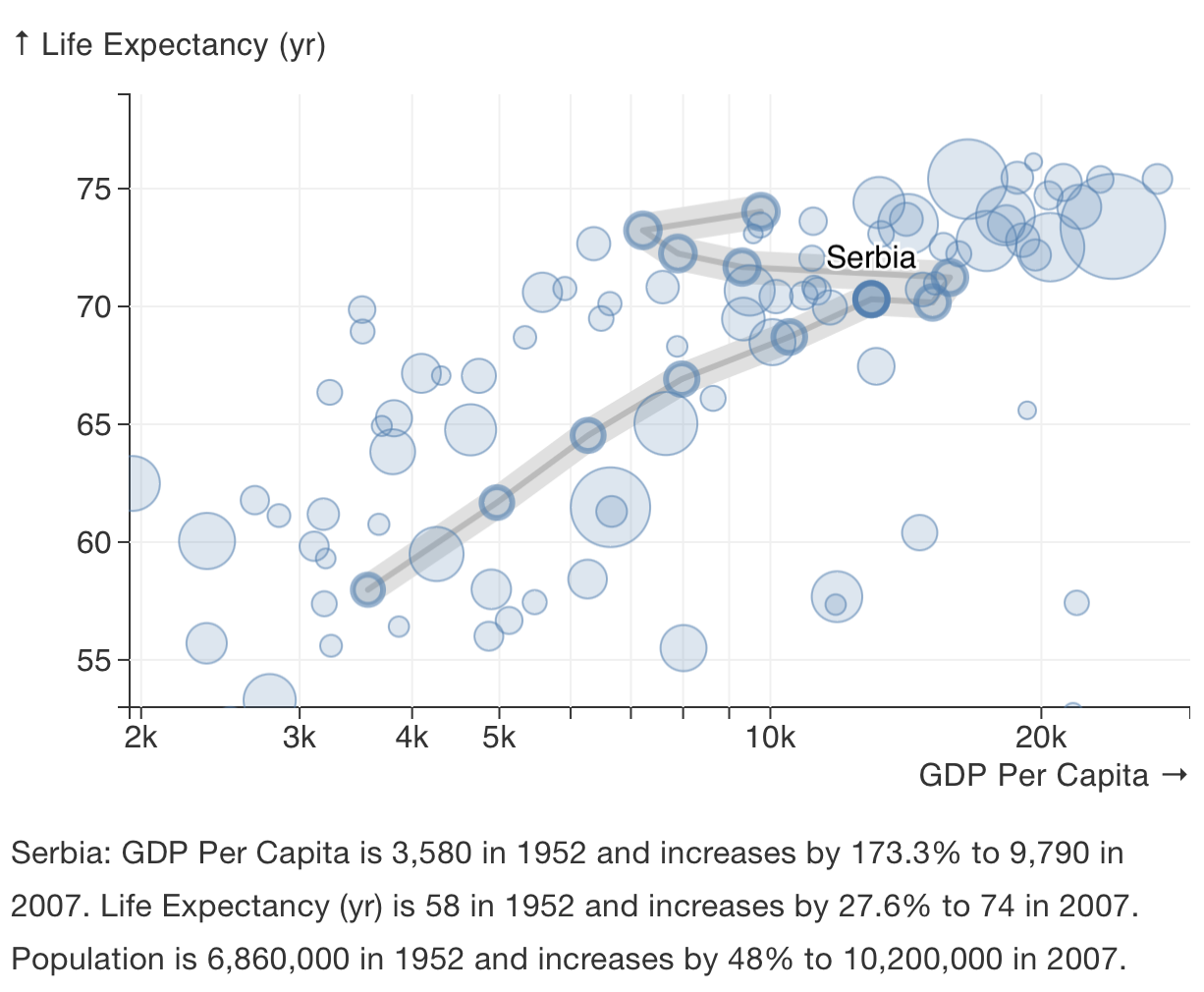}
    \caption{Visualizing global economies from the Gapminder dataset~\cite{gapminder_gapminder_nodate}. The user can navigate the chart using multiple modalities and see or hear text descriptions of each mark. The interactive version is one of the examples at \url{https://dig.cmu.edu/counterpoint}.}
    \label{fig:gapminder}
\end{figure}

Next, we develop a version of the classic animated Gapminder chart~\cite{gapminder_gapminder_nodate} to show Counterpoint's ability to complement existing frameworks and simplify addressing accessibility considerations.
This chart, shown in \cref{fig:gapminder}, consists of standard SVG-based axes and a Canvas-based bubble chart, as well as variable-width lines that animate in when the user hovers on or selects a country.
Bubbles and lines are represented as \texttt{Mark}s whose spatial attributes react to an animatable \texttt{year} attribute.

While rendering this chart in Canvas provides greater control and performance than SVG, it would be difficult to make such a chart responsive and accessible for people with disabilities using existing tools.
For users with vestibular or photosensitive disabilities, we configure marks to listen to the \texttt{RenderContext}'s \texttt{prefersReducedMotion} property, triggering a redraw when the user changes their system-wide motion preference. 
With reduced motion enabled, we replace line-draw and motion animations with fade transitions (which we achieve by adding and removing copies of each mark from the stage).
Furthermore, for screen reader users we implement keyboard and touch-based navigation as well as automatic high- and low-level descriptions using Data Navigator~\cite{elavsky_data_2024}.
We use the marks for each country to generate a navigable structure that selects each country and shows a description of its data values as the user presses arrow keys or swipes through the visualization. 
We also summarize the country's overall movement trajectory for a higher level description.
Counterpoint helps reduce the amount of code needed to maintain state upon adding these alternative interaction modes.

\section{Discussion}

We presented Counterpoint, a framework that helps manage complex visualization state and incurs minimal performance overhead while enabling the development of feature-rich systems.
Beyond the use cases presented, Counterpoint's architecture makes it suitable for a wide variety of large-scale animated visualizations, including network visualizations, flow charts, choropleths, beeswarm plots, and others.
Abstracting state management can allow developers to focus on incorporating important features such as progressive rendering, support for multiple input modalities, and accessibility for people with visual disabilities.
These affordances help bridge the gap between the power of scalable Web graphics APIs and the intricacies of engineering usable visualizations.

Like many state management frameworks, Counterpoint is targeted towards highly complex and custom software systems, so its utility for more standard use cases is limited.
While implementing simple animated charts is possible and straightforward with Counterpoint, it provides the most benefit when a plot needs thousands of animating marks and multiple forms of interaction and animation.
As a future direction, incorporating Counterpoint into existing visualization authoring tools could help developers easily leverage its functionality regardless of plot complexity.

Future work can build on this initial release by incorporating additional transitions and interpolation functions that support common rendering use cases.
For example, Counterpoint could facilitate the creation of parametric and sequenced animations such as the mark type transitions developed in DynaVis~\cite{heer_animated_2007}.
Counterpoint's API could also be augmented with simple shortcuts for common application needs, such as configuring staging with fade transitions or live-mapping data elements to marks.
We are releasing Counterpoint as open-source (version 1.0.4 as of this publication), and we invite the data visualization community to contribute feedback and improvements.


\acknowledgments{%
  We thank Eli Kranjec, Donghao Ren, Halden Lin, Yannick Assogba, and Fred Hohman for feedback and contributions to the framework design. This work was supported by a National Science Foundation Graduate Research Fellowship (DGE2140739), and by the Carnegie Mellon University Center of Machine Learning and Health.
}

\bibliographystyle{abbrv-doi-hyperref}

\bibliography{references}








\end{document}